\documentclass{aastex631}

\shorttitle{Formation of Counter-rotating Stellar Disks}
\shortauthors{Bao et al.}
\graphicspath{{./}{figures/}}

\begin{document}

\title{Different Formation Scenarios of Counter-rotating Stellar Disks in Nearby Galaxies}

\author{Min Bao}
\affiliation{Department of Astronomy, Nanjing University, Nanjing 210093, China \\
Key Laboratory of Modern Astronomy and Astrophysics (Nanjing University), Ministry of Education, Nanjing 210093, China \\
Collaborative Innovation Center of Modern Astronomy and Space Exploration, Nanjing 210093, China}
\affiliation{School of Physics and Technology, Nanjing Normal University, Nanjing 210023, China}

\author{Yanmei Chen}
\affiliation{Department of Astronomy, Nanjing University, Nanjing 210093, China \\
Key Laboratory of Modern Astronomy and Astrophysics (Nanjing University), Ministry of Education, Nanjing 210093, China \\
Collaborative Innovation Center of Modern Astronomy and Space Exploration, Nanjing 210093, China}

\correspondingauthor{Yanmei Chen}
\email{chenym@nju.edu.cn}

\author{Pengpei Zhu}
\affiliation{Department of Astronomy, Nanjing University, Nanjing 210093, China \\
Key Laboratory of Modern Astronomy and Astrophysics (Nanjing University), Ministry of Education, Nanjing 210093, China \\
Collaborative Innovation Center of Modern Astronomy and Space Exploration, Nanjing 210093, China}
\affiliation{Department of Astronomy and Astrophysics, University of California, Santa Cruz, CA 95064}

\author{Yong Shi}
\affiliation{Department of Astronomy, Nanjing University, Nanjing 210093, China \\
Key Laboratory of Modern Astronomy and Astrophysics (Nanjing University), Ministry of Education, Nanjing 210093, China \\
Collaborative Innovation Center of Modern Astronomy and Space Exploration, Nanjing 210093, China}

\author{Dmitry Bizyaev}
\affiliation{Apache Point Observatory and New Mexico State University, P.O.  Box 59, Sunspot, NM,  88349-0059,  USA}
\affiliation{Sternberg Astronomical Institute, Moscow State University, Moscow, Russia}

\author{Ling Zhu}
\affiliation{Shanghai Astronomical Observatory, Chinese Academy of Sciences, 80 Nandan Road, Shanghai 200030, China}

\author{Meng Yang}
\affiliation{Shanghai Astronomical Observatory, Chinese Academy of Sciences, 80 Nandan Road, Shanghai 200030, China}

\author{Minje Beom}
\affiliation{Department of Astronomy, New Mexico State University, Las Cruces, NM 88001, USA}

\author{Joel R. Brownstein}
\affiliation{Department of Physics and Astronomy, University of Utah, Salt Lake City, UT 84112, USA}

\author{Richard R. Lane}
\affiliation{Centro de Investigaci\'{o}n en Astronom\'{i}a, Universidad Bernardo O'Higgins, Avenida Viel 1497, Santiago, Chile}

\begin{abstract}
Using the integral field unit (IFU) data from Mapping Nearby Galaxies at Apache Point Observatory (MaNGA) survey, we select a sample of 101 galaxies with counter-rotating stellar disks and regularly-rotating ionized gas disk. We classify the 101 galaxies into four types based on the features of their stellar velocity fields. The relative fractions and stellar population age radial gradients of the four types are different in blue cloud (BC), green valley (GV) and red sequence (RS) populations. We suggest different formation scenarios for the counter-rotating stellar disks, and the key factors in the formation of counter-rotating stellar disks include: (1) the abundance of pre-existing gas in progenitor, (2) the efficiency in angular momentum consumption.
\end{abstract}
 

\section{Introduction} \label{sec:intro}

The so-called 2$\sigma$ galaxies have two peaks along the major axes in the stellar velocity dispersion fields. A natural explanation is that they host two counter-rotating stellar disks \citep{2014ASPC..486...51C}. Counter-rotating disks could form during gas-rich disk-disk mergers \citep{2001Ap&SS.276..909P, 2009MNRAS.393.1255C}, or as a result of external gas accretion or gas-rich minor mergers with angular momentum misaligned with the pre-existing disk followed by star formation \citep{1998ApJ...506...93T, 2017MNRAS.471.1892B, 2021MNRAS.500.3870K}. These different formation mechanisms will result in different properties of the two disks, including luminosity fraction, stellar population age and metallicity. The kinematic properties and metallicity of gas in these galaxies provide additional information of their formation scenario.

Since the first discovery of counter-rotating stellar disks in NGC 4550 \citep{1992ApJ...394L...9R}, several cases have been studied over the years, e.g., NGC 5719 \citep{2011MNRAS.412L.113C}, NGC 3593 \citep{2013A&A...549A...3C}, NGC 4138 \citep{2014A&A...570A..79P}, NGC 448 \citep{2016MNRAS.461.2068K}, NGC 5102 \citep{2017MNRAS.464.4789M} and IC719 \citep{2018A&A...616A..22P}. With high quality spectra, the absorption lines in these star-star counter-rotators are featured with explicit double-peaks. Thus, the two disks could be decomposed through spectra fitting, e.g., \cite{2011MNRAS.412L.113C, 2016MNRAS.461.2068K}. With spectral decomposition method, the surface brightness, stellar velocity, stellar velocity dispersion, age and metallicity of each disk could be obtained. It turned out that in all these individual cases, the stellar disk which is co-rotating with the gas component is younger with lower metallicity, indicating that (i) the progenitor obtains counter-rotating gas from external; (ii) the newly formed stars inherit the angular momentum from the counter-rotating gas, leading to formation of two counter-rotating stellar disks. \cite{2021MNRAS.tmp.3448B} selected $\sim$53 galaxies with counter-rotating stellar disks from about 4000 galaxies in MaNGA DR16, finding that in most cases, the gas co-rotates with the younger disk, which is consistent with studies of individual cases. In addition, \cite{2021MNRAS.tmp.3448B} separated their sample into three categories: (1) 11 cases with ongoing star formation; (2) 25 galaxies with multimodality in the stellar populations; (3) galaxies with unimodality in the stellar populations. They suggested that these different types of galaxies with counter-rotating stellar disks have different formation scenarios, but without discussions on what these scenarios are.

In this work, we select a sample of 101 galaxies with two counter-rotating stellar disks from the MaNGA Product Launch 10 (MPL10, \citealt{2015ApJ...798....7B, 2015AJ....149...77D, 2015AJ....150...19L, 2016AJ....152...83L}). The sample selection is displayed in Section \ref{sec:sample}. We classify galaxies with two counter-rotating stellar disks into four types in section \ref{sec:results} and discuss their formation scenarios in section \ref{sec:conclusions}.

\section{Sample Selection} \label{sec:sample}

The data products in this work are drawn from the MaNGA MPL10 \citep{2017AJ....154...86W, 2016AJ....151....8Y, 2016AJ....152..197Y}, which includes 9456 unique galaxies. The global $r$-band effective radius ($Re$) defined as the radius which includes half light of a galaxy, is adopted from NASA-Sloan Atlas (NSA, \citealt{2011AJ....142...31B}). The MaNGA observation covers at least to 1.5 $Re$ for most of the targeted galaxies. The spatially resolved properties, i.e., stellar/gas velocity, stellar velocity dispersion and Lick index 4000\AA\ break ($D_{n}4000$) which traces the stellar population age, are obtained from the MaNGA data analysis pipeline (DAP, \citealt{2019AJ....158..231W}). The gas velocity is traced by ionized hydrogen (i.e., H$\alpha$), while all the emission line centers are tied together in the velocity space in MaNGA DAP. We match the MPL10 sample with the literature catalog \citep{2015ApJS..219....8C} to obtain the global stellar mass ($M_{\star}$) and star formation rate (SFR) for 8431 out of 9456 galaxies, and estimate the specific star formation rate as sSFR $\equiv$ SFR / $M_{\star}$.

\begin{figure}[ht!]
    \epsscale{1}
    \plotone{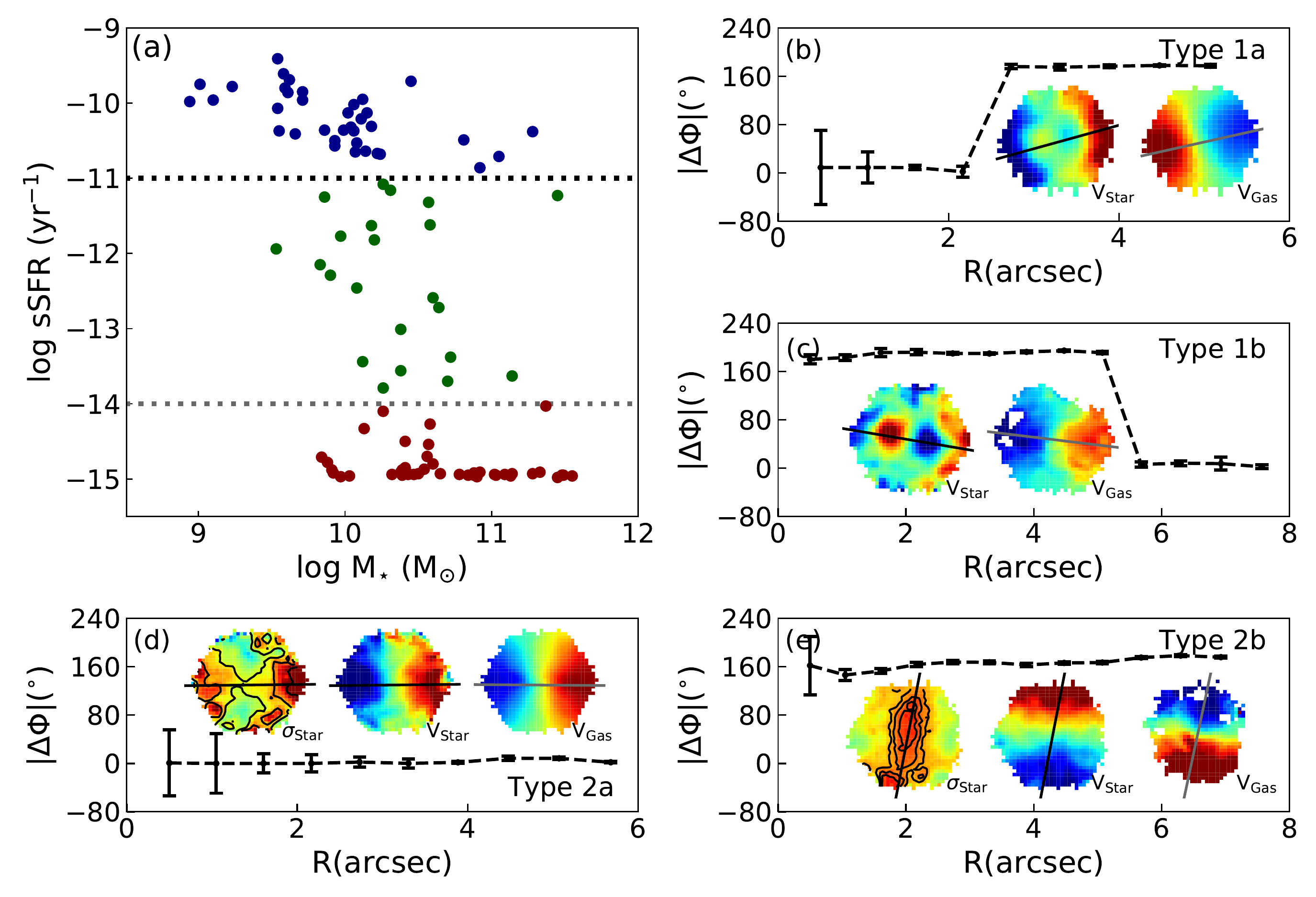}
    \caption{(a): Specific star formation rate as a function of stellar mass. The black and grey dotted lines separate galaxies into BC (blue), RS (red) and GV in-between them (green). (b) (c) (d) (e): The difference of kinematic position angles between stellar and gas components along radius (black dashed lines), as well as stellar and gas velocity fields for Types 1a, 1b, 2a and 2b CRDs. The black and grey solid lines mark the stellar and gas kinematic major axes. The stellar velocity dispersion fields for Types 2a and 2b are also displayed in panels (d) and (e). The black contours indicate the $\sigma$ peaks along the stellar major axes. \label{fig1}}
    \end{figure}

2$\sigma$ galaxy is introduced by \cite{2011MNRAS.416.1680C} to describe the feature of off-center but symmetric peaks in the stellar velocity dispersion map along the major axis of a galaxy. Basically, it origins from galaxy with two counter-rotating stellar disks. Considering the prominance of bulge and the modest spatial resolution of MaNGA, the appearance of one feature does not necessarily imply the other one. Thus, we select galaxies with two counter-rotating stellar disks according to one of the following criteria:

\begin{itemize}
    \item [1.]
    Search the stellar velocity dispersion fields for two $\sigma$ peaks or single elongate $\sigma$ peak along the major axis: (1) calculate the mean velocity dispersions along major and minor axes; (2) seek galaxies where the velocity dispersion along the major axis is higher than minor axis, we get 421 candidates; (3) visually inspect the stellar velocity dispersion fields of the 421 candidates to make sure there are two $\sigma$ peaks (see stellar velocity dispersion map in Figure \ref{fig1}d) or single elongate $\sigma$ peak (see stellar velocity dispersion map in Figure \ref{fig1}e) along the major axis. 99 galaxies are selected.

    \item [2.]
    Search the stellar velocity fields for counter-rotation between inner and outer disks: (1) use KINEMETRY package \citep{2006MNRAS.366..787K} to fit the stellar rotation in each 0.5arcsec (pixel size) ellipse and obtain the position angle; (2) calculate the position angle differences ($|\Delta \phi|$) between the nearby ellipses, searching for the galaxies which satisfy $|\Delta \phi| > 150^{\circ}$, we get 453 candidates; (3) visually inspect the stellar velocity fields of the 453 candidates for distinguishable counter-rotating stellar disks (see the stellar velocity fields in Figures \ref{fig1}b and \ref{fig1}c). 36 galaxies are selected.
 
\end{itemize}

There are 17 overlaps between galaxies selected according to the first and second criteria. Other 19 galaxies with counter-rotating stellar disks but without off-center $\sigma$ peaks show low line-of-sight velocities, which can be explained by a low inclination angle \citep{2021A&A...654A..30R} or a low intrinsic rotation velocity. In total, we have 118 candidates with counter-rotating stellar disks. A few galaxies with counter-rotating stellar disks don't show ionized-gas emission \citep{2011MNRAS.414.2923K, 2016MNRAS.461.2068K}. To clearly assess the gas angular momentum direction and compare it with stars, we exclude 17 galaxies where the gas emission is too weak to be detected or the gas rotation doesn't have regular pattern. The final sample applied in the following study consists of 101 galaxies with counter-rotating stellar disks and regularly-rotating ionized gas disk. We will refer them as CRDs in the following sections. Figure \ref{fig1} shows the sSFR as a function of $M_{\star}$ for the 101 CRDs. We separate the 101 galaxies into three populations: 36 blue cloud (BC) galaxies with $\log$(sSFR) $>$ -11 yr$^{-1}$ (blue filled circles in Figure \ref{fig1}a), 43 red sequence (RS) galaxies with $\log$(sSFR) $<$ -14 yr$^{-1}$ (red filled circles in Figure \ref{fig1}a) and another 22 green valley (GV) galaxies in-between the two extremes (green filled circles in Figure \ref{fig1}a). The spatially resolved BPT diagrams \citep{2006MNRAS.372..961K} of the 43 RS CRDs show that most of them are dominated by extended low-ionization emission-line regions, i.e., extended LIERs \citep{2016MNRAS.461.3111B}.

\section{Different Types of Counter-rotating Stellar Disks} \label{sec:results}
We classify the sample galaxies into two branches. For the first branch, the counter-rotating stellar disks are clearly distinguishable in the stellar velocity fields. This branch can be divided into two types: Types 1a and 1b. Figure \ref{fig1}b shows an example of Type 1a. The black dashed line shows the difference of position angles ($|\Delta\Phi|$) between stars and gas along radius. It is clear the gas disk is co-rotating with the inner stellar disk and counter-rotating with the outer one. Figure \ref{fig1}c shows an example of Type 1b where the gas disk is co-rotating with the outer stellar disk. For the second branch, one of the stellar disks dominates the stellar velocity field, no evidence of two counter-rotating stellar disks can be found from the velocity fields, but the two $\sigma$ peaks or single elongate $\sigma$ peak along the major axis is clearly observed in the stellar velocity dispersion fields. This branch can also be divided into two types: Types 2a and 2b. Figures \ref{fig1}d and \ref{fig1}e show examples of Types 2a and 2b, respectively. The difference of position angle between stars and gas indicates star-gas co-rotation in Type 2a and counter-rotation in Type 2b. \cite{2014ASPC..486...51C} summarized the individual galaxies with counter-rotating stellar disks from the earlier literatures, and all the cases with regularly-rotating ionized gas disk can be included in our classification. \cite{2006ApJ...649L..79M} found ring structure in NGC 7742, where the gas is counter-rotating with stars. We visually inspect the SDSS $g, r, i$ color images and H$\alpha$ flux maps of CRDs and find only three galaxies showing ring structures.

\begin{figure}[ht!]
    \epsscale{1}
    \plotone{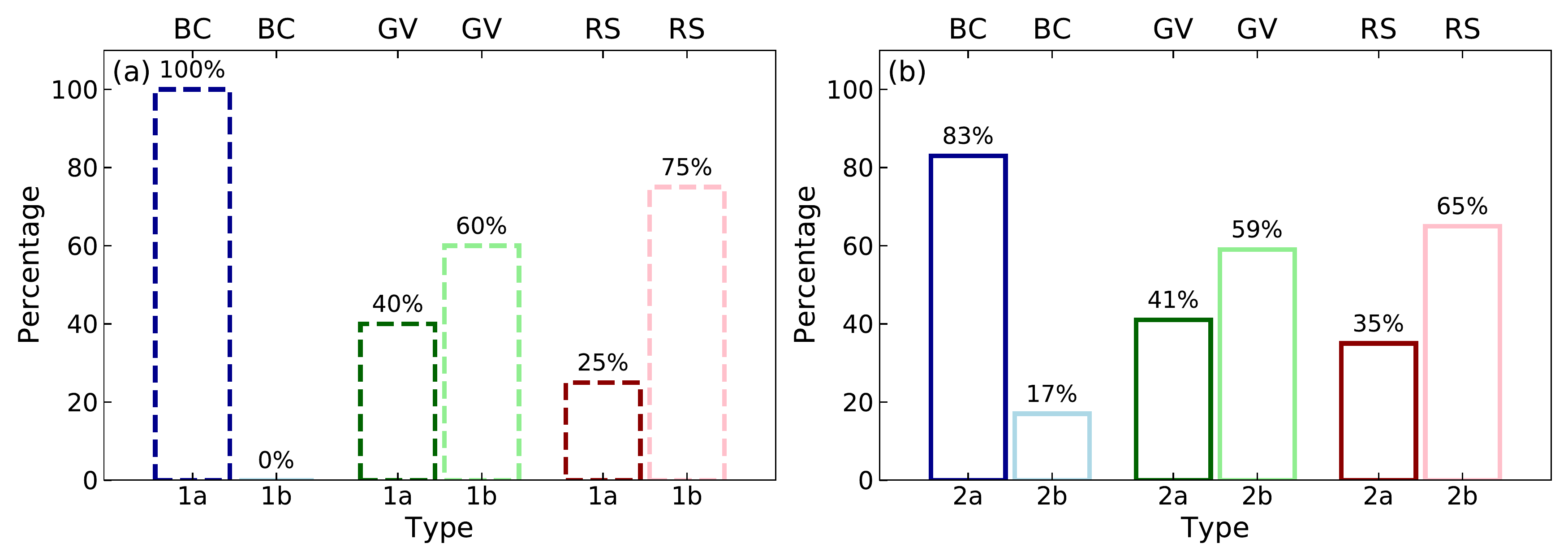}
    \caption{(a): The relative contributions of Types 1a and 1b. Blue, green and red histograms represent BC, GV and RS galaxies. The dark-color dashed-line histograms show the fractions of Type 1a (the gas disk co-rotates with the inner stellar disk). The light-color dashed-line histograms show the fractions of Type 1b (the gas disk co-rotates with the outer stellar disk). (b): The relative contributions of Types 2a and 2b. Blue, green and red histograms represent BC, GV and RS galaxies. The dark-color solid-line histograms show the fractions of Type 2a (star-gas co-rotators). The light-color solid-line histograms show the fractions of Type 2b (star-gas counter-rotators). The percentage of each type is marked on top of the histogram. \label{fig2}}
    \end{figure}

Figure \ref{fig2}a shows the relative contributions of Types 1a and 1b in BC (blue), GV (green), QS (red) galaxies, while the relative contributions of Types 2a and 2b are in Figure \ref{fig2}b. From Figure \ref{fig2}a, we find that 100\% BC galaxies are classified as Type 1a (the gas disk co-rotates with the inner stellar disk) and this fraction decreases monotonically from BC to 25\% in RS galaxies. In contrast, the fraction of Type 1b (the gas disk co-rotates with the outer stellar disk) increases from BC to RS galaxies. Similarly, in Figure \ref{fig2}b, 83\% CRDs in BC are classified as Type 2a (star-gas co-rotators), this percentage falls to 41\% in GV galaxies and to 35\% in RS galaxies. While the percentage of Type 2b (star-gas counter-rotators) increases from BC to GV and to RS. 

\begin{figure}[ht!]
    \epsscale{1}
    \plotone{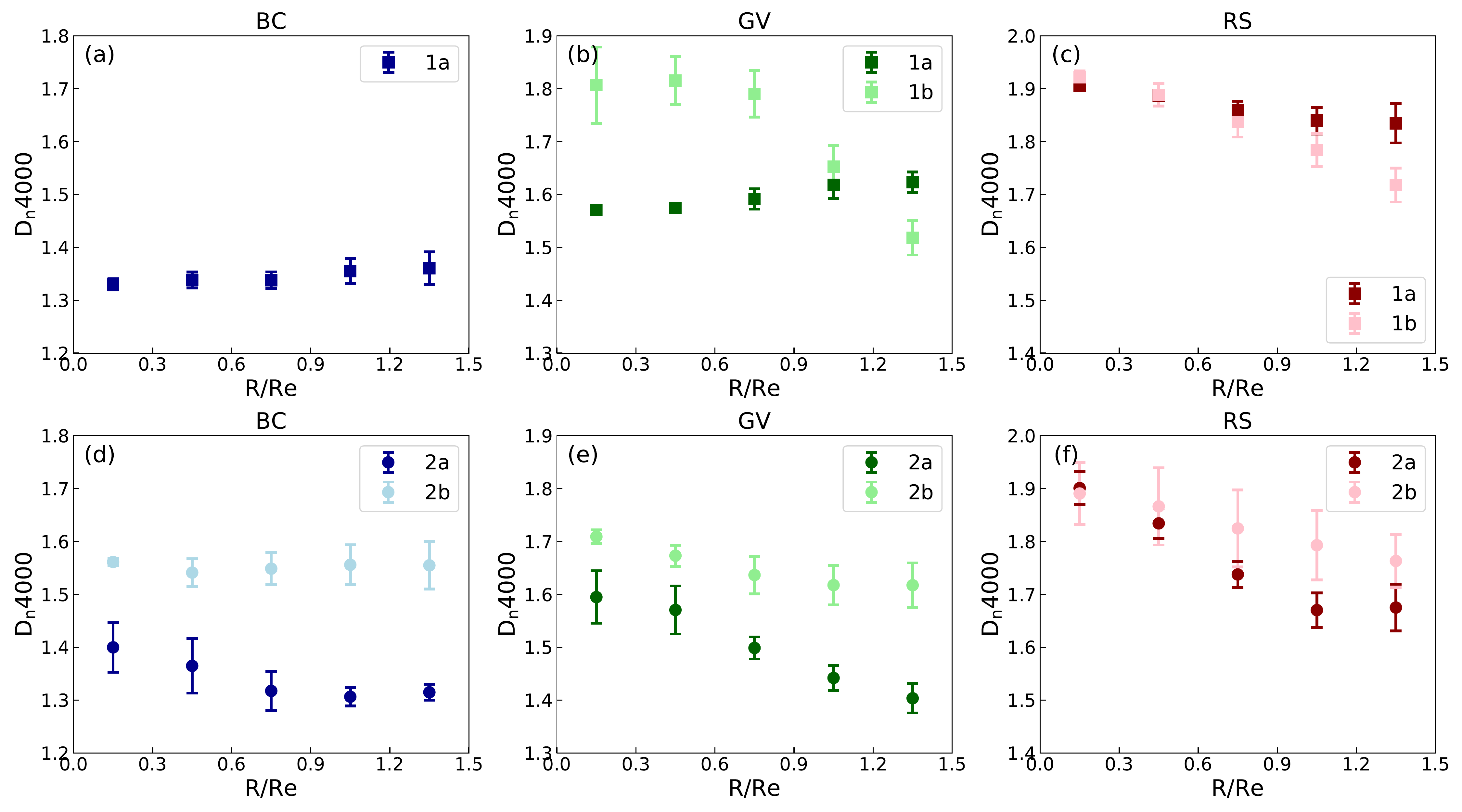}
    \caption{$D_{n}4000$ radial gradients in different types of CRDs. Blue, green and red colors correspond to the BC, GV and RS galaxies. (a) (b) (c): The dark-color squares mark the median values of $D_{n}4000$ in each radial bin of Type 1a. The light-color squares mark the median values of $D_{n}4000$ in each radial bin of Type 1b. (d) (e) (f): The dark-color circles mark the median values of $D_{n}4000$ in each radial bin of Type 2a. The light-color circles mark the median values of $D_{n}4000$ in each radial bin of Type 2b. The error bars show the 40\% to 60\% percentiles of the distribution. \label{fig3}}
    \end{figure}

Figure \ref{fig3} shows the $D_{n}4000$ radial gradients in different types of CRDs. The top row for Type 1 and the bottom row for Type 2. BC, GV and RS galaxies are shown from left to right columns, respectively. The blue, green and red squares/circles mark the median values of $D_{n}4000$ in each radial bin for BC, GV and RS, and the error bars show the 40\% to 60\% percentiles of the distribution. Considering $D_{n}4000$ is a good indicator of the light-weighted stellar population age, Figure \ref{fig3}a shows Type 1a CRDs in BC have young stellar population across the whole galaxy with slightly positive gradient, indicating the inner disk is slightly younger than the outer disk \citep{2013A&A...549A...3C, 2017MNRAS.464.4789M}.  In Figure \ref{fig3}b, the gradients in $D_{n}4000$ for Type 1a and Type 1b GV galaxies are totally different, with a positive gradient for Type 1a and negative for Type 1b, indicating Type 1a has younger inner disk than the outer disk, while the inner disk is older than the outer disk in Type 1b. Although both Types 1a and 1b in RS galaxies (Figure \ref{fig3}c) show negative gradients in $D_{n}4000$, indicating the inner disk is older than the outer disk, the slopes of the gradients are clearly different between Types 1a and 1b. Type 1a has a slightly younger (older) stellar population in the central (outer) region than Type 1b. In Figures \ref{fig3}d, \ref{fig3}e and \ref{fig3}f, negative gradients dominate in Types 2a and 2b \citep{2015A&A...581A..65C, 2018A&A...616A..22P} except a flat gradient of Type 2b in BC galaxies. It is obvious Type 2a has younger stellar population than Type 2b in BC, GV and RS galaxies.

\section{Formation Scenarios of Counter-rotating Stellar Disks} \label{sec:conclusions}

\begin{figure}[ht!]
    \epsscale{1}
    \plotone{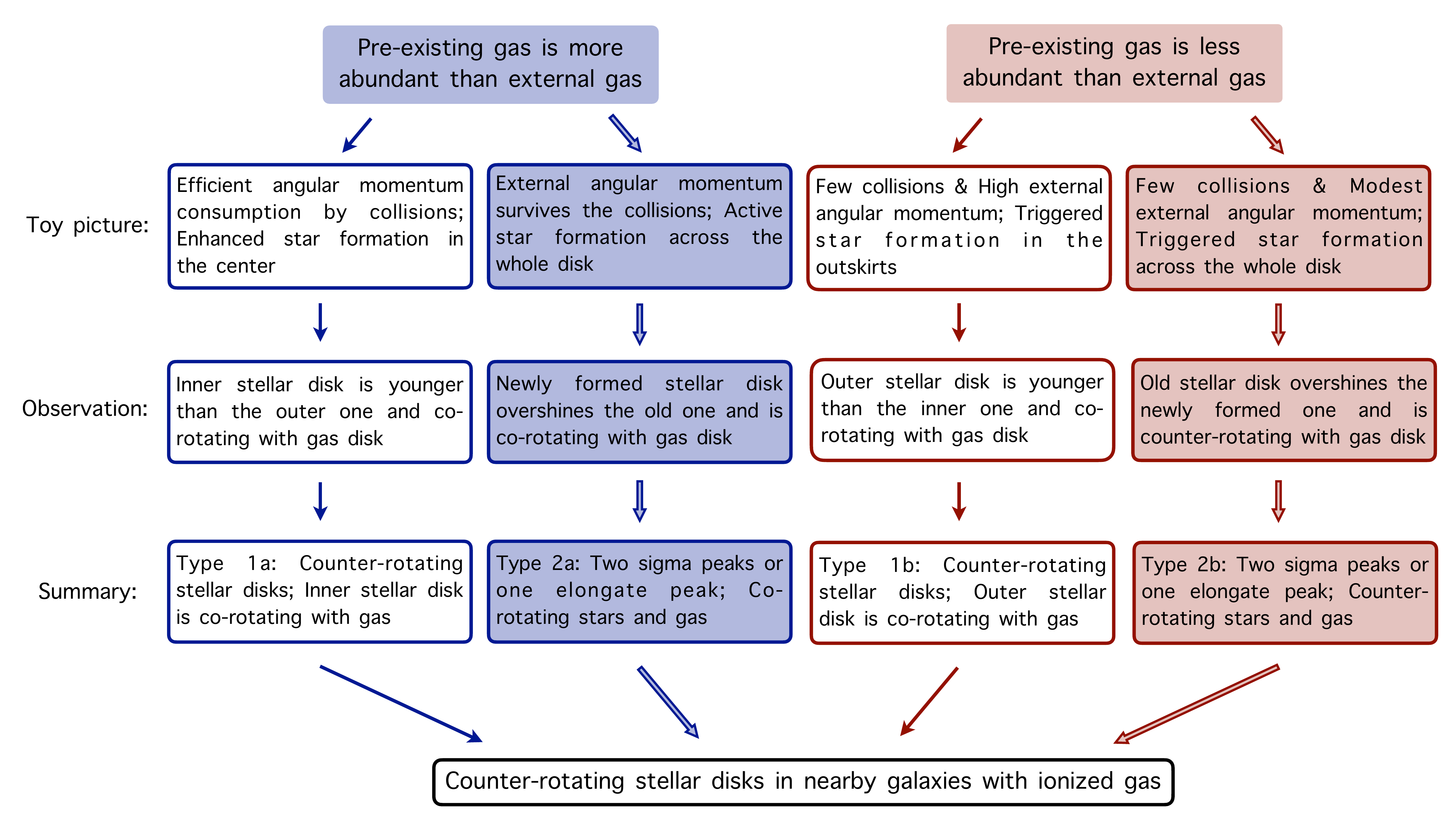}
    \caption{Formation scenarios of counter-rotating stellar disks in the nearby galaxies with regularly-rotating ionized gas disks. \label{fig4}}
\end{figure}

In this work, we select a sample of 101 galaxies with counter-rotating stellar disks and classify them into four types based on the features of their stellar velocity fields. As shown in Figures \ref{fig2} and \ref{fig3}, the relative fractions of four types and $D_{n}4000$ radial gradients are different in BC, GV and RS galaxies, which indicate different formation scenarios of CRDs. We summarize the primary formation scenarios for the four types in Figure \ref{fig4}.

The formation scenarios of Types 1a and 2a CRDs are shown in the first and second columns in Figure \ref{fig4}. In these two types, the gas in progenitors is more abundant than the accreted gas \citep{2004A&A...424..447P}. The collisions between pre-existing and external gas efficiently consume the angular momentum in Type 1a, leading gas inflow and triggering star formation in the central region. The enhanced star formation in the central region can explain the observations that the inner stellar disk is younger than the outer one in Type 1a BC and GV galaxies (Figures \ref{fig3}a and \ref{fig3}b), also Type 1a has slightly younger stellar population in the central region than Type 1b in RS galaxies (Figure \ref{fig3}c). In this case, the inner disk is composed of newly formed stars, which is younger than the outer disk. And the newly formed stars are co-rotating with the gas where they formed from. Part of the angular momentum of external gas survives the gas-gas collisions in Type 2a, and the newly formed stars spread over the whole disk, overshing the old one. Since the newly formed stars inherit the angular momentum from gas, gas-star co-rotation is observed. The monotonically decreasing trends of Types 1a and 2a fractions from BC to GV and to RS shown in Figure \ref{fig2} also support the idea that the more abundant pre-existing gas, the higher efficiency of angular momentum consumption, and easier to form Types 1a and 2a CRDs.

The formation scenarios of Types 1b and 2b CRDs are shown in the third and fourth columns in Figure \ref{fig4}. The pre-existing gas is less abundant than the accreted gas, which results in the inefficient angular momentum consumption. Thus star formation of Type 1b is triggered in the outskirts, leading to a younger outer disk, which is co-rotating with the gas. The younger stellar population in the outskirts of Type 1b than Type 1a in GV and RS galaxies (Figures \ref{fig3}b and \ref{fig3}c) can be explained in this scenario. In Type 2b, the newly formed stars spread over the whole disk, but the star formation is less active than Type 2a. The old stellar disk, which is counter-rotating with the accreted gas, overshines the newly formed one, so that Type 2b has older stellar population than Type 2a in BC, GV and RS galaxies (Figures \ref{fig3}d, \ref{fig3}e and \ref{fig3}f). In Figure \ref{fig2}, the high fractions of Types 1b and 2b in RS galaxies support the idea that poor pre-existing gas is preferred in these types of CRDs.

\begin{acknowledgments}
    Y. C acknowledges support from the National Key R\&D Program of China (No. 2017YFA0402700), the National Natural Science Foundation of China (NSFC grants 11733002, 11922302), the China Manned Space Project with NO. CMS-CSST-2021-A05. M. B acknowledges support from the National Natural Science Foundation of China (No. 11873032). The authors are very grateful to the referee for valuable comments and suggestions which improved the work.

    Funding for the Sloan Digital Sky Survey IV has been provided by the Alfred P. Sloan Foundation, the U.S. Department of Energy Office of Science, and the Participating Institutions. 

    SDSS-IV acknowledges support and resources from the Center for High Performance Computing  at the University of Utah. The SDSS website is www.sdss.org. SDSS-IV is managed by the Astrophysical Research Consortium for the Participating Institutions of the SDSS Collaboration including the Brazilian Participation Group, the Carnegie Institution for Science, Carnegie Mellon University, Center for Astrophysics | Harvard \& Smithsonian, the Chilean Participation Group, the French Participation Group, Instituto de Astrof\'isica de Canarias, The Johns Hopkins University, Kavli Institute for the Physics and Mathematics of the Universe (IPMU) / University of Tokyo, the Korean Participation Group, Lawrence Berkeley National Laboratory, Leibniz Institut f\"ur Astrophysik Potsdam (AIP),  Max-Planck-Institut f\"ur Astronomie (MPIA Heidelberg), Max-Planck-Institut f\"ur Astrophysik (MPA Garching), Max-Planck-Institut f\"ur Extraterrestrische Physik (MPE), National Astronomical Observatories of China, New Mexico State University, New York University, University of Notre Dame, Observat\'ario Nacional / MCTI, The Ohio State University, Pennsylvania State University, Shanghai Astronomical Observatory, United Kingdom Participation Group, Universidad Nacional Aut\'onoma de M\'exico, University of Arizona, University of Colorado Boulder, University of Oxford, University of Portsmouth, University of Utah, University of Virginia, University of Washington, University of Wisconsin, Vanderbilt University, and Yale University.
\end{acknowledgments}

\end{document}